\def\arxiv{1}
\def\removeforpaper{1}
\newcommand{\revFIVE}[1]{#1}

\newcommand{\optLevelONEternary}[2]{
\ifdefined\optLevelONE
#2
\else
#1
\fi
}
\def\numauthors{3}

\ifdefined\arxiv\fi
\ifdefined\longversion
\newcommand{\longtext}[1]{#1}
\else
\newcommand{\longtext}[1]{}
\fi
\ifdefined\arxiv
\documentclass[conference]{IEEEtran}
\else
\documentclass[sigconf]{acmart}
\fi

\ifdefined\arxiv 
\usepackage{cite}
\usepackage{amssymb}
\usepackage{amsmath}
\usepackage{booktabs}
\usepackage{textcomp}
\title{Extending the RISC-V ISA for Efficient RNN-based 5G Radio Resource Management}

\author{\IEEEauthorblockN{Renzo Andri\IEEEauthorrefmark{1},
Tomas Henriksson\IEEEauthorrefmark{3} and
Luca Benini\IEEEauthorrefmark{1}\IEEEauthorrefmark{2}}
\IEEEauthorblockA{\IEEEauthorrefmark{1}Integrated Systems Laboratory, 
ETH Z\"urich,
Zurich, Switzerland}
\IEEEauthorblockA{\IEEEauthorrefmark{2}DEI, University of Bologna, Bologna, Italy}\IEEEauthorblockA{\IEEEauthorrefmark{3}Huawei Technologies Sweden AB, Kista, Sweden}}

\else

\fi

\usepackage{pgffor}

\newcommand{\rnnfigures}[1]{./RNN/figures/#1}
\newcommand{\rnntables}[1]{./RNN/tables/#1}
\usepackage{multicol}
\usepackage{multirow}
\usepackage{lipsum}
\usepackage{fp}
            \usepackage{algorithm}
            \usepackage{csvsimple}
\usepackage{algorithmic}
\usepackage{array}
\usepackage{siunitx}
\usepackage{tikz}
\usepackage{pgfplots}
\usepackage{pgf-pie}
 \usetikzlibrary{shadows, positioning, calc}
\pgfplotsset{compat=1.13}
\definecolor{color0}{RGB}{255,0,0}
\definecolor{color1}{RGB}{0,255,0}
\definecolor{color2}{RGB}{102,140,217}
\definecolor{color3}{RGB}{16,150,24}
\definecolor{color4}{RGB}{153,0,153}
\definecolor{color5}{RGB}{255,165,0}
\definecolor{cieee0}{HTML}{00629b}
\definecolor{cieee1}{RGB}{255,199,44}
\definecolor{cieee2}{RGB}{232,119,34}
\definecolor{cieee3}{RGB}{186,12,47}
\definecolor{cieee4}{RGB}{119, 37, 131}
\definecolor{cieee5}{RGB}{120, 190, 32}
\definecolor{cieee6}{RGB}{0, 132, 61}
\definecolor{cieee7}{RGB}{0,  159,  223}
\definecolor{colorintable}{HTML}{108AC7}
\DeclareMathAlphabet{\mathcal}{OMS}{cmsy}{m}{n}
\SetMathAlphabet{\mathcal}{bold}{OMS}{cmsy}{b}{n}
\ifdefined\printversion
\newcommand{\todo}[1]{}
\newcommand{\todolater}[1]{\textcolor{red}{\textbf{#1}}}
\else
\usepackage{todonotes}
\newcommand{\todolater}[1]{}
\fi
\newcommand{\calc}[2]{\FPeval{\calcresult}{round(#2,#1)}\calcresult}
\ifdefined\arxiv
\usepackage{hyperref}
\else
\hypersetup{draft} 
\fi
\newcommand{\x}{$\times$}

\begin{document}


\newcommand{\titleFIR}{Extending the RISC-V ISA}
\newcommand{\titleSEC}{for Efficient RNN-based 5G Radio Resource Management}
\newcommand{\shortTitle}{\titleFIR\ \titleSEC}
\newcommand{\myvec}[1]{\ensuremath{\begin{pmatrix}#1\end{pmatrix}}}

\ifdefined\arxiv \maketitle 
\bibliographystyle{IEEEtran} \else \title[\shortTitle]{\titleFIR\\\titleSEC}\fi
\ifdefined\arxiv \else
\author{Renzo Andri}
\affiliation{%
  \institution{Integrated Systems Lab., ETH Zurich}
  \streetaddress{Gloriastrasse 35}
  \city{Zurich}
  \country{Switzerland}
  \postcode{8092}
}
\email{andrire@iis.ee.ethz.ch}


\author{Tomas Henriksson}
\affiliation{%
  \institution{Huawei Technologies Sweden AB,}
  \streetaddress{}
  \city{Kista}
  \country{Sweden}}
\email{Tomas.Henriksson@huawei.com}

\author{Luca Benini}

\affiliation{%
  \institution{Integrated Systems Lab., ETH Zurich}
  \streetaddress{Gloriastrasse 35}
  \city{Zurich}
  \country{Switzerland}
  \postcode{8092}
}
\affiliation{%
  \institution{University of Bologna}
  \streetaddress{Gloriastrasse 35}
  \city{Bologna}
  \country{Italy}
  \postcode{}
}
\email{benini@iis.ee.ethz.ch}
\fi

\begin{abstract} 

Radio Resource Management (RRM) in 5G mobile communication is a challenging problem for which Recurrent Neural Networks (RNN) have shown promising results. Accelerating the compute-intensive RNN inference is therefore of utmost importance. Programmable solutions are desirable for effective 5G-RRM top cope with the rapidly evolving landscape of RNN variations. In this paper, we investigate RNN inference acceleration by tuning both the instruction set and micro-architecture of a micro-controller-class open-source RISC-V core. We couple HW extensions with software optimizations to achieve an overall improvement in throughput and energy efficiency of 15$\times$ and 10$\times$ w.r.t. the baseline core on a wide range of RNNs used in various RRM tasks.\ifdefined\anonymous
\footnote{Hardware, Software and benchmarks have been open sourced on GitHub \mbox{\url{https://github.com/anonymoususer/anonymousrepo}}}
\else
\footnote{Hardware, software and benchmarks have been open sourced on GitHub \mbox{\url{https://github.com/andrire/RNNASIP}}}
\fi
\end{abstract}

\ifdefined\arxiv \else


\fi

\ifdefined\arxiv \else\maketitle\fi

\section{Introduction}  
\revFIVE{Radio Resource Management is challenging as it aims at achieving maximum utilization of the limited publicly available frequency bands \cite{tripathi2006radio}, under highly heterogeneous traffic (e.g., tiny sensor-nodes vs. mobile routers), and rapidly varying radio signal propagation conditions. Notably, RRM tasks have to be executed in the frame of milliseconds, which exclude compute-intensive algorithms \cite{Sun2017}. Presently, 5G applications impose strict new intensive requirements on radio communication systems: 1) very high reliability and low-latency for autonomous vehicles, 2) very high bandwidth requirements for video telephony and virtual reality, and 3) massive machine-to-machine communication for the Internet of Things. These challenging requirements ask for extending the existing cellular network with more antennas, improving antenna efficiency, and more effective RRM. Therefore, more advanced allocation algorithms are required to distribute limited resources (e.g., frequency bands, transmit power, data rates) to mobile clients efficiently.}



Typically, RRM problems have been modeled with full observability and solving convex problems with traditional optimization approaches. Exhaustive search methods led to very high computation costs \cite{Ahmed2018}, and sub-optimal solutions based on Lagrangian relaxation, iterative distribution optimization, and other heuristic approaches had convergence issues and lacked guarantees \cite{Ahmed2018}. Traditional methods like the weighted sum-rate MSE algorithm \cite{shi2011iteratively} and fractional programming \cite{naeem2013optimal} are iterative, and most of them need to perform complex operations (e.g., matrix inversion or SVD) in every single iteration. It is, therefore, extremely challenging to push these methods to the throughput and scale required for 5G-RRM. Recently, neural networks have gained increasing attention for 5G-RRM. At the physical layer, RNNs have been used to compensate for imperfections and nonlinearities and collision detection in the RF domain \cite{Yao2019, Yao2019ArtificialNetworks}. This is getting even more important for high-frequency communication, where absorption starts to strongly depend on the environment, and for ultra-dense cell networks where cross-tier interference has to be compensated \cite{ge20165g}. Classic multi-layer perceptron \cite{ye2018deep, Ghadimi2017, Yu2017, Nasir2018}, (recurrent) Long Short-Term Memories LSTM \cite{Challita2017, Naparstek2017}, and Convolution Neural Networks \cite{Lee2018} have been used at the data-link layer, which is responsible for resource allocation, including dynamic resource scheduling of frequency bands, dynamic range, and handover control. Reinforcement learning-based deep Q-Learning networks \cite{mnih2015human} have been used for several typical RRM problems like dynamic spectrum access utilization \cite{Naparstek2017,ye2018deep,wang2018deep}, power level selection \cite{ye2018deep, Ghadimi2017, Nasir2018}, rate control \cite{Ghadimi2017}, and time-slotted optimization \cite{Yu2017}.

These networks are less computationally demanding than classical RRM algorithms, but they are far from trivial. Specialized and efficient stand-alone Neural Networks accelerators have been presented recently \cite{reuther2019survey}. Nevertheless, hardwired RNN accelerators cannot cope with the flexibility requirements found in a typical RRM setting, as base stations typically stay in the field for a very long time, while RRM algorithms are rapidly evolving. FPGA-based acceleration has been explored for RNN inference to retain flexibility. For instance, LSTM acceleration on FPGA achieving up to 13\,GMAC/s/W, has been presented in Cao et al.  \cite{Cao2019} and Gao et al.\cite{gao2018}. To further increase efficiency, compression techniques (e.g., block-circulant weight matrices, pruning with zero-skipping \cite{gao2018, Cao2019}) have been applied, and a top (effective) energy efficiency of 82\,GMAC/s/W on a Xilinx Zynq-7100 FPGA has been presented \cite{gao2018}. Nevertheless, these compression schemes have not yet been proven to work for the networks used in the RRM field, and FPGAs have a cost envelope that is not compatible with massive and dense deployment, as required in 5G networks. To address this intertwined flexibility, efficiency, and cost challenges, we propose to enhance the open and royalty-free RISC\nobreakdash-V ISA and leverage the availability of high-quality open-source cores based on this widely supported ISA.
We demonstrate a micro-controller class RISC-V core with RNN-enhancements for RRM acceleration, and we couple hardware extensions with software optimizations. We achieve an energy efficiency of 218\,GMAC/s/W, and a throughput of 566\,MMAC/s, which is an improvement of 10\x{} and 15\x, respectively, over the baseline open-source core. Such an order-of-magnitude boost is obtained thanks to data reuse with output feature map tiling (1.9\x), adding custom activation instructions (13\% within LSTMs), merging load and compute (1.15\x/1.7\x), and input FM tiling (5\%).

The proposed extensions maintain backward compatibility with the baseline RISC-V ISA and have a very small overhead (3.4\%) in area and no increase in the longest path. Improvements are consistently achieved over a quite diverse set of RNNs used for various RRM tasks, thereby confirming the flexibility of our approach.

\if False
\cite{Ahmed2018}
\begin{itemize}
    \item complexity and heterogeneity of emerging 5G cellular networks
    \item resource allocation move to the node 
    \item previously optimization theory used (but complex wireless channels, heterogeneous systems, diverse QoS user requirements
    \item previously used exhaustive search methods, genetic algorithms, combinatorial $\rightarrow$ high computational complexity
    \item sub-optimal solution based on techniques such as Lagrangian relaxation, iterative distribution optimization, heuristic algorithms, and auction/game theory may also be computationally intensive and can be far from optimal. problem with convergence and knowledge of confidence is tricky
    \item a communication system suffers from hardware and transmission channel impairments including inter-modulation and amplifier distortion, quantization loss, quadrature imbalance, multi-path fading, interference, path-loss. 
    \item problem of optimal utilization (antenna, channels, bands, beams, codes, bandwidths, 
\end{itemize}
AI Def. 5G Radio Access Networks \cite{Yao2019ArtificialNetworks}.
\begin{itemize}
    \item Massive multiple-input multiple-output antenna systems, millimeter-wave communications, and ultra-dense networks are the key enablers to allow for the development and deployment of 5G systems.
    \item 5G through aggressively increased spectral efficiency and channel bandwidth and higher density.
    \item The International Telecommunication Union classifies 5G into enhanced mobile broadband (eMBB), massive machine-type communication (mMTC), and ultra-reliable and low latency communication (URLLC).
    \begin{itemize}
        \item eMBB: massive video streaming and virtual/augmented reality
        \item mMTC massive sensing, monitoring, metering in support of the massive deployment of IoT
        \item URLLC latency-sensitive services including autonomous driving, drones, and the tactile internet (transmitting a physical sense of touch remotely)
    \end{itemize}
    \item for cognitive link adaptation, resource scheduling, signal classification, and carrier sensing/collision detection
    \item RNN, able to mitigate the imperfections and nonlinearities of RF components
    \item upper communication layers task like network optimization and resource management more with DBN and CNN.
    \item Massive Multiple-input multiple-output systems are perceived as leading candidates for 5G, but high number of base stations require a large number of power amplifiers. $\rightarrow$ design trade-off between linearity and energy efficiency and cost, but emerging energy/ and spectrum-efficient wideband wireless communication systems are vulnerable to non-linear distortion.
    \item next-generation communication systems will operate in a more dynamic environment and different bands; the dynamic range requirements are becoming more demanding $\rightarrow$ stricter linearity specifications needed.
    \item ANN is able to tackle non-linear problems at the PHY layer, including power amplifiers nonlinearity tracking, pre-distortion, and impairment correction.
    \item Need of high data-rates, asks for high-frequency communication systems, but not very reliable due to physical interaction (e.g., absorption) $\rightarrow$  use of low and high frequencies (need smart scheduling and cooperation).
    \item densification (ultra-dense small cells) ref8, finer-grained resource allocation, and cross-tier interference mitigation. The radio network prediction, evolution, and optimization can be made by active learning based on network coverage capacity and quality maps. The AI agent helps with radio resource management, interference management, and handover control.
    \end{itemize}
    
    arm helium: M-profile Vector Extension MVE), goals: (8.1)
    - significantly enhance microcontroller dsp and ml performance
    - retain interrupt and latency guarantees of arm cortex-m
    - integrate with system-wide secutiry provided by Arm TrustZone
    - Simplify DSP and ML development within the ARM ecosystem toolchains
    - fit within the ppa requirements of microcontroller systems
    
    8 instead of 16 vector registers (vs neon)
    down to 8-bit (vs. neon)
    complex number operation SIMD
    
    128-bit Q registers (4xFP32 SIMD)
    -register count unchanged (same area)
    -unlike neon, can target general-purpose accumulators
    
    128-bit vectors =>
    - 16x8-bit integers
    - 8x INT16 or FP16
    - 4x INT32 or FP32
    2x64-bit accumulators INT64
    32-bit accumulators FP32
    
    single code for high performance/low area, different #beat (number of operations on the simd executed)
    hardware loops
    VMLA mac accumulated in scalar register file (reducing vector register pressure), supports signed/unsigned 8,16,32 bit inputs 
    
    interleaved load/store post-increment
    
\fi
\section{Related Works}
\subsection{ML Compute Platforms}
With the machine learning revolution, a variety of different ML compute platforms have been presented in industry and academia, spanning from high-performance server accelerators (e.g., Google's TPU cores) to embedded platforms (e.g., Nvidia Jetson Xavier) to stand-alone application-specific accelerators \cite{reuther2019survey}. We are not aware of any RNN acceleration engine targeting  RRM applications. General-purpose processors have been extended with new matrix and vector extensions to handle the common compute patterns in Neural Networks. In the Advanced Vector Extensions AVC-512 of the x86 ISA, Intel added the \texttt{VNNIW} instruction extension, which include 16$\times$32-bit SIMD vector operation for efficient convolution kernels in single-precision float FP16 and accumulations in double-precision float FP32 and since Cascade Lake (2019) the fixed-point version (\texttt{VNNI}) with 8-bit (e.g., \texttt{VPDBUSD}) and 16-bit (e.g., \texttt{VPDBUSSD}) vector product with 32-bit accumulation \cite{IntelCorp.2019IntelArchitectureInstructionReference}. The AARCH64 Neon extensions in the ARMv8-A processor series provides special SIMD instructions for sum-dot-products (e.g., \texttt{BFDOT}) and 2$\times$2 matrix-matrix multiplications (e.g., \texttt{BFMMLA}) with 2-way SIMD in brain floating-point format \texttt{bfloat16}. Recently, ARM presented the M-profile Vector Extensions MVE (Helium) for their embedded processor family Cortex-M. Helium instructions feature computations in various SIMD-formats (INT8/16/32, FP16/32), hardware loops, interleaved post-increment load/stores \cite{armhelium}. However, Intel typically focuses on the high-performance high-cost processor market, and the Helium extensions are not yet available in HW implementations. 

Besides ISA extension, also highly-optimized SW kernels have been developed exploiting these instructions. These include utilizing parallel SIMD computations (e.g., 16-bit \cite{lai2018cmsis}, 8-bit \cite{garofalo2019pulp}) and data reuse with appropriate tiling. Tiling helps to reduce data loads from memory and reuse data with the local registerfile. Output FM tiling (OFM), where several outputs are calculated in parallel and input FM loads can be shared, has been commonly used (e.g., \cite{lai2018cmsis, garofalo2019pulp}). Furthermore, convolutional layers can be reformulated as matrix-matrix multiplications with the im2col technique \cite{in2col}. This allows to tile both the input and output FM spatially in $m\times n$-sized tiles and thus reduces the number of loads from $\mathcal{O}(mn)$ to $\mathcal{O}(m+n)$, as both weights and input FM pixels can be reused. Previous work has mainly focused on on CNNs \cite{lai2018cmsis, garofalo2019pulp}. Still, this two-dimensional tiling cannot be applied to (non-convolutional) LSTMs and Linear Layers, which are the main network kernels used in RRM applications.

\ifdefined\notdefined
\begin{itemize}
    \item Advanced Vector Extensions AVX extension of the x86 ISA \cite{IntelCorp.2019IntelArchitectureInstructionReference}
    \item VNNIW 32-bit float VP4DPWSSD(S) and accumulated in 64-bit float (Knights Mill)
    \item VNNI, designed to accelerate convolutional neural network-based algorithm (introduced in 2019) with Cascade Lake and Ice Lake 
    \item 8-bit VPDBUSD and 16-bit VPDBUSSD and accumulated in 32-bit (fixed-point)
    \item ARM8 introduces Neon and SVE vector instructions using Bfloat (FP 8,7)\cite{Stephens2019DevelopmentsArmv8.6-A}
    \item AArch64 Neon instructions: 
    \begin{itemize}
        \item BFDOT: 2 word BF dotproduct with FP32 result
        \item BFMMLA: computes 2x2 elements at a time of a matrix multiplication
        \item BFMLAL: simple product either odd or even element and accuml in FP32
        \item BFCVT: converts FP32 to BF16
    \end{itemize}
    \item Neural Network accelerator for GPU \cite{yazdanbakhsh2015neural}
    \item elimintating fetch/decoding during execution, reducing accesses to the memory/register file by storing the parameters and partial results in small buffers within the SIMD lanes, implementing sigmoid as a lookup table.
    \item SparCE: Sparsity Aware Genereal-Purpose Core Extnesions to Acceleratre Deep Neural Networks\cite{sen2018sparce}
    \begin{itemize}
        \item instructions are skipped before fetched
        \item evaluated on 4-way SIMD ARMv8 processor using the OpenBLAS library (8-15\% improvement of execution time) on the cycle-accurate gem5 architectural simulator 
        \item if load is zero skip operation
    \end{itemize}
\end{itemize}
\fi
Neural Networks are commonly trained in floating-point format. Still, recently, it has been shown that integer-aware training allows us to use more energy and area efficient fixed-point without any significant accuracy drop, especially 16-bit quantization \cite{lin2016fixed}, but even eight and fewer bits \cite{jacob2018quantization}.

Finally, RNNs use transcendental activation functions, which are computationally complex. Previously, there have been 4 approaches to accelerate the computation of these functions: piecewise linear approximation (PLA)\cite{lai2018cmsis}, low-order Taylor series expansion (e.g., 2nd order \cite{lin2008digital}), LUT with adaptive value granularity \cite{leboeuf2008high}, or a small neural network \cite{tsai2015hardware}. We use a PLA approach, but differently from previous work, we exploit the symmetry property of \texttt{tanh} and \texttt{sig}, we take into account fixed-point quantization and evaluate in detail the error introduced by different numbers of interpolation intervals, rather than selecting a high number of intervals (i.e., 128 in ARM's CMSIS-NN \cite{lai2018cmsis}).







\subsection{RISC-V and RI5CY}
The RISC-V ISA \cite{waterman2014risc} has recently become the de facto standard in open-source and free instruction set architecture. RISC-V provides plenty of encoding space for extensions and is therefore suitable for application-driven processor customization while maintaining compatibility with the baseline ISA. In this work, we rely on the RI5CY \cite{Gautschi2017}, a high-quality, silicon-proven, and open-source core supporting the standard RISC-V RV32IMFC ISA (including integer, integer multiplications, single-precision floating-point, and compressed instructions). Additionally, RI5CY supports the Xpulp ISA extensions featuring extended fixed-point support (e.g., on-the-fly re-quantization and saturation), SIMD instructions, post-increment store and loads, and hardware loops. 


\subsection{Benchmark Suite and Neural Networks}
We have selected an application benchmark consisting of 10 neural networks which have been presented recently in the RRM domain. These networks differ in network types (Fully-Connected Neural Layers (\cite{ye2018deep, Sun2017,  Yu2017, Eisen, Nasir2018, Ahmed2018, wang2018deep}, Long-short Term Memories \cite{Challita2017, Naparstek2017}, Convolutional Neural Network \cite{Lee2018}), learning methods (Supervised \cite{Challita2017, Sun2017, Eisen, Lee2018}, reinforcement-based \cite{Naparstek2017, ye2018deep,  Yu2017, Nasir2018, wang2018deep}, unsupervised \cite{Ahmed2018}), application (cellular networks \cite{Challita2017, Ahmed2018}, peer-to-peer communication \cite{Naparstek2017}, wireless communication systems \cite{Yu2017,Eisen,Nasir2018,Lee2018, wang2018deep}, wired communication \cite{Sun2017}), and optimization metric (throughput \cite{Challita2017,Naparstek2017,Sun2017,Yu2017,Eisen,Nasir2018,Ahmed2018,Lee2018,wang2018deep}, fairness \cite{Challita2017, Naparstek2017}, latency \cite{ye2018deep}, energy efficiency \cite{ Lee2018}). A detailed description of the networks can be found in the project report \cite{huawei1}. Three main ML kernels are used within these networks: Fully-connected layers (or Multi-Layer Perceptron MLP), Long-short Term Memories LSTM, and Convolutional Neural Network CNN Layer. A fully-connected layer connects all input (neurons) $\mathbf{x}\in \mathbb{R}^m$ to all outputs (neurons) $\mathbf{o}\in \mathbb{R}^n$ and is described with the following matrix-vector multiplication and the corresponding weight matrix  $\mathbf{W} \in \mathbb{R}^{n\times m}$: $\mathbf{o} = \mathbf{b} +\mathbf{W}\mathbf{x}$. 
LSTM are recurrent networks able to learn time series and are described by $m$ input neurons and $n$ internal memory cells $c_t$, $n$ hidden states $h_t$ and the corresponding matrix-vector multiplications, point-wise vector-vector multiplications/additions (i.e., Hadamard product $\mathbf{a}\circ \mathbf{b}= (a_i\cdot b_i)_i$), and point-wise application of sigmoid and hyperbolic tangent activation functions:
\begin{align}
\mathbf{o_t}(x_t, h_{t-1}) &= \text{sig}(\mathbf{W_{o}} \mathbf{x_t} + \mathbf{U_{o}} \mathbf{h_{t-1}} + \mathbf{b_o}) \\
\mathbf{f_t}(x_t, h_{t-1}) &= \text{sig}(\mathbf{W_{f}} \mathbf{x_t} + \mathbf{U_{f}} \mathbf{h_{t-1}} + \mathbf{b_f}) \\
\mathbf{i_t}(x_t, h_{t-1}) &= \text{sig}(\mathbf{W_{i}} \mathbf{x_t} + \mathbf{U_{i}} \mathbf{h_{t-1}} + \mathbf{b_i}) \\
\mathbf{g_t}(x_t, h_{t-1}) &= \text{tanh}(\mathbf{W_{c}} \mathbf{x_t} + \mathbf{U_{c}} \mathbf{h_{t-1}} + \mathbf{b_c}) \\
\mathbf{c_t}(x_t, c_{t-1}) &= \mathbf{f_t} \circ \mathbf{c_{t-1}} + \mathbf{i_t} \circ \mathbf{g_t}  \\
\mathbf{h_t}(o_t, c_{t}) &= \mathbf{o_t} \circ \text{tanh}(\mathbf{c_t)}
\end{align} Whereas the weight matrices $W_{o}, W_{f}, W_{i}, W_{g} \in \mathbb{R}^{n\times m}$ and $ U_{o},$ $U_{f}, U_{i}, \allowbreak U_{c} \in \mathbb{R}^{m\times m}$ and bias vectors $b_o,b_f,b_i,b_c\in\mathbb{R}^{m}$.

Finally, CNN layers exploit the translation invariance in the data (e.g., in images) and map $n$ $h_{im, in}\times w_{im, in}$-sized input channels $i_n\in \mathbb{R}^{h_{im, in}\times w_{im, in}}$ $k$ $h_{im, out}\times w_{im, out}$-sized output channel maps by applying $h_{k}\times b_{k}$-sized convolution filters $w_{k,n}\in  \mathbb{R}^{h_{k}\times b_{k}}$ to every input channel for every output channel.
\ifdefined\optLevelONE
\vspace{-1mm}
 \begin{align}
\mathbf{o_k} &= \mathbf{C_k}+\sum_{n\in I}\underbrace{{\mathbf{i_n} \ast \mathbf{w_{k,n}}}}_{\mathbf{\tilde{o}_{k,n}}} 
\label{eq:1}
\end{align}
\else
\[
\mathbf{o_k} = \mathbf{C_k}+\sum_{n\in I}\underbrace{{\mathbf{i_n} \ast \mathbf{w_{k,n}}}}_{\mathbf{\tilde{o}_{k,n}}} \] \[
  = C_k+\sum_{n\in I}\underbrace{\Bigg(\sum^{b_k-1}_{a=0}\sum^{h_k-1}_{b=0}{i_n\left(x+a,y+b\right) \cdot w_{k,n}(a,b)}\Bigg)}_{\tilde{o}_{k,n}(x,y)}
\]
\fi



\ifdefined\detailed
\unsure{add detailed descriptions of the networks}
\fi
\section{HW/SW Extension and Optimizations}
\begin{table*}[h!]
\small
\centering
\caption{Cycle and Instruction Count Optimizations for the entire RRM benchmark suite (\textbf{RISCY in bold}, \textcolor{black}{\textbf{new ext. in blue}})}

\ifdefined\arxiv \def\scalefactor{0.92} \else \vspace{-3mm}\def\scalefactor{1}\fi
\scalebox{\scalefactor}{
\begin{tabular}{@{}rrr@{}p{2mm}@{}rrr@{}p{2mm}@{}rrr@{}p{2mm}@{}rrr@{}p{2mm}@{}rrr@{}p{2mm}@{}} \toprule
\multicolumn{3}{c}{\textbf{a) w/o opt (RV32IMC)}} &&\multicolumn{3}{c}{\textbf{b) +SIMD/HWL (Xpulp)}} &&\multicolumn{3}{c}{\textbf{c) +Out-FM Tile./tanh/sig}} &&\multicolumn{3}{c}{\textbf{d) +pl.sdotsp instruction}}&&\multicolumn{3}{c}{\textbf{e) +Input FM Tiling}} \\ \cmidrule{1-3}\cmidrule{5-7}\cmidrule{9-11}\cmidrule{13-15}\cmidrule{17-19}
Instr. & \textit{k}cycles  & \textit{k}instrs    &&    Instr. & \textit{k}cyc.  & \textit{k}instrs   & &      Instr. & \textit{k}cyc.  & \textit{k}instrs   &&     Instr. & \textit{k}cyc.&  \textit{k}instrs  &&     Instr. & \textit{k}cyc.&  \textit{k}instrs \\ \midrule
   addi  &3'269 &3'269&& \textbf{lw!}& 2'432& 1'621        &     &\textbf{lw!} & 894&  893 && \textcolor{colorintable}{\textbf{pl.sdot}} & 811&  811  & &\textcolor{colorintable}{\textbf{pl.sdot}} & 817&  817        \\  
   bltu  &3'248 &1'627&& \textbf{pv.sdot}&  811&    811      &   & \textbf{pv.sdot} & 811&  811  && \textbf{lw!} & 166&  83 && \textbf{lw!} & 83&  83        \\  
   lh &3'248 &3'248&& addi&   22&     22      &      &lw &   9&    9  && lw &   9&    9 &  &         lw &   39&    35     \\  
   sw &1'627 &1'627&& jal&   10&      5       &    & sw &   8&    8 && sw &   8&    8 &   &          sw &   16&    16  \\  
 lw &1'627 &1'627&&    sh&   10&     10      &    & add &   7&    6 && add &   7&    6 &   &         d.srai &   8&   8    \\  
 mac &1'621 &1'621&&    srai&   10&     10      &  &    \textcolor{colorintable}{\textbf{tanh,sig}} &   0.4&    0.4 && \textcolor{colorintable}{\textbf{tanh,sig}} &   0.4&    0.4 &  &     \textcolor{colorintable}{\textbf{tanh,sig}} &   0.4&    0.4      \\ \cmidrule{1-3}\cmidrule{5-7}\cmidrule{9-11}\cmidrule{13-15}\cmidrule{17-19}
 oth. &   43 &   32 &&    oth.&   28&     27      &  &     oth. &   26&    26 && oth. &   30&    29 &   &       oth. &   17&    10     \\  \midrule
 $\Sigma$&14'683&13'051& &   $\Sigma$& 3'323&  2'506 &  &       $\Sigma$& 1'756&  1'753 & & $\Sigma$ &1'028& 943 && $\Sigma$ &980& 969    \\ \cmidrule{1-3}\cmidrule{5-7}\cmidrule{9-11}\cmidrule{13-15}\cmidrule{17-19}
 Impr.&\multicolumn{2}{c}{Baseline (1\x)}&&Impr.&\multicolumn{2}{c}{\calc{1}{14683/3323}\x}&&Impr.&\multicolumn{2}{c}{\calc{1}{14683/1756}\x{} (\calc{1}{3323/1756}\x)}&&Impr.&\multicolumn{2}{c}{\calc{1}{14683/1028}\x{} (\calc{1}{1756/1028}\x)}&&Impr.&\multicolumn{2}{c}{\calc{1}{14683/980}\x{} (\calc{2}{1028/980}\x)}    \\ \bottomrule
\end{tabular}}
 \label{tab:rnninstructioncomparison}\ifdefined\optLevelONE \vspace{-0.3cm}\fi
\end{table*}

\subsection{Baseline Implementation (SW)}
We have developed a straight-forward implementation (e.g., organizing matrix-vector multiplication as a double nested loop over all inputs and outputs) of all required network kernels in C where weights and data values are encoded in 16-bit fixed-point format (i.e, $Q_{3.12}$). This format offers a good compromise between accuracy/robustness and energy-efficiency/throughput, and most importantly, does not require fixed-point aware retraining that would be necessary for smaller bit-widths. The C implementation is compiled with standard GCC 7.1.1 for RISC-V RV32IMFC ISA and was run on the RI5CY core. The instruction count for the entire benchmark suite is shown in Tab.~\ref{tab:rnninstructioncomparison}a and is used as the baseline for further comparisons. 

\subsection{SIMD, HWL and post-increment load (HW)}
As a first optimization step, we re-wrote the code to exploit Xpulp extensions as much as possible. The 16-bit data (weights and inputs) are packed into the packed SIMD vector format (i.e., \texttt{v2s}), allowing the compiler to map every two subsequent input FM $p(2c_i)$ and $p(2c_i+1)$ and the corresponding weights $(c_o, 2c_i)$ and $w(c_o, 2c_i+1)$ to a \texttt{mac} using a single \texttt{pv.sdotsp.h} instruction without the need of custom intrinsics.
\begin{align}
    o(c_o)       &= b(c_{o}) + \sum_{c_i=0}^{c_{in}/2} \myvec{p(2c_i)\\p(2c_i+1)}\myvec{w(c_o, 2c_i)\\w(c_o, 2c_i+1)}
\end{align}
The next optimization is to reduce the overhead of loop control instructions in small loop bodies that are seen in such operations by using hardware loops that are part of the Xpulp extensions. 
The hardware loop does not use any additional instructions during the loop execution, but requires loop index manipulation instructions (i.e., \texttt{pl.setup}) to set three registers: a loop counter (\texttt{rB}), the loop start \texttt{PC+4} and the loop end (\texttt{PC+rA}). When the \texttt{PC} reaches the loop end, the controller decrements the loop counter and jumps back to the loop start until the loop counter reaches zero.

The final optimization is to take advantage of post-increment load-word instruction (i.e., \texttt{lw!}) to increment the data pointer for weights and input feature maps at the same time as executing the load word instruction, saving a separate \texttt{addi} instruction in the process. Combining these three techniques results in  \calc{1}{14683/3323}$\times$ reduction w.r.t. to the unmodified RISC-V IMC baseline in the number of instructions executed as can be seen in  Tab.~\ref{tab:rnninstructioncomparison}b.


\subsection{Output Feature Map Tiling (SW)}

To compute one MAC two loads to the memory are needed: one for the weight and one for the value of the corresponding input neuron. Fortunately, the read for the input value can be reused for several outputs. The output features are therefore organized in tiles of $N$ output channels and the contribution of the input neurons is calculated for all output neurons of the current input neuron. These partial sums can be stored in registers and are not written back to the memory until all input activations have been weighted and accumulated. Algorithm\,\ref{alg:outputtilingscheme} gives an overview of the implementation and scheduling of the output FM tiling. 
\begin{algorithm}[h!]
  \caption{Fully-Connected Layer with Output FM Tiling}
  \label{alg:outputtilingscheme}
\begin{algorithmic}[1]
\REQUIRE{All weights $w_{mn}$ and input activations $i_m$ for all input channels $m\in c_{in}$ and output channels $n\in c_{out}$ in memory}

\FORALL{d-sized output channel tiles $\tilde{o}_k=\{o_{k\cdot d}, ..., o_{(k+1)\cdot d}\}$}
    \FORALL{output channels $o_l$ in $\tilde{o}_k$}
        \STATE temp\_out[$o_l$] = 0\;
    \ENDFOR
    \FORALL{input channels $i_l\in c_{in}$}
        \STATE temp\_in=Mem($i_l$)\;
        \STATE \#unroll following loop\;
        \FORALL{output channel $o_k$ in tile $\tilde{c}_{out}$}
            \STATE w=Mem($w_{o_k, i_l}$)\;
            \STATE temp\_out[$o_k$] $\mathrel{{+}{=}}$ temp\_in * w\;
        \ENDFOR
    \FORALL{output channels $o_k$ in $\tilde{o}_k$}
        \STATE temp\_out[$o_k$] = temp\_out[$o_k$] $>>$ 12 // requantize\;
        \STATE Mem($o_k$) = temp\_out[$o_k$] \;
    \ENDFOR
    \ENDFOR
\ENDFOR

\end{algorithmic}
\end{algorithm}
The load of one input FM can thus be shared by $N$ \texttt{pl.sdotsp} instructions (executing 2 MAC operations on 16-bit operands), and thus just $\mathcal{O}(1+1/N)$ loads are needed per compute operation. $N$ can be increased until the available registers are exhausted, and data has to be pushed onto the stack memory; furthermore, the load latency can be hidden by the compiler by rearranging the instructions. Previous work has shown that the tiling can be extended to the feature-level in case of a convolutional layer if the input feature map is rearranged and replicated (i.e., in2col) such that the convolution becomes a matrix-matrix multiplication \cite{lai2018cmsis, garofalo2019pulp}. 

In this paper, we focus mainly on the optimizations for LSTMs and MLPs, as these network kernels are mostly used in the selected RRM benchmark suite and have not been discussed in previous work. As can be seen in Tab.\,\ref{tab:rnninstructioncomparison}c, the optimal tiling brings an additional improvement of 1.89$\times$ on the RRM benchmark.

\begin{figure*}[h!]
\centering
\input{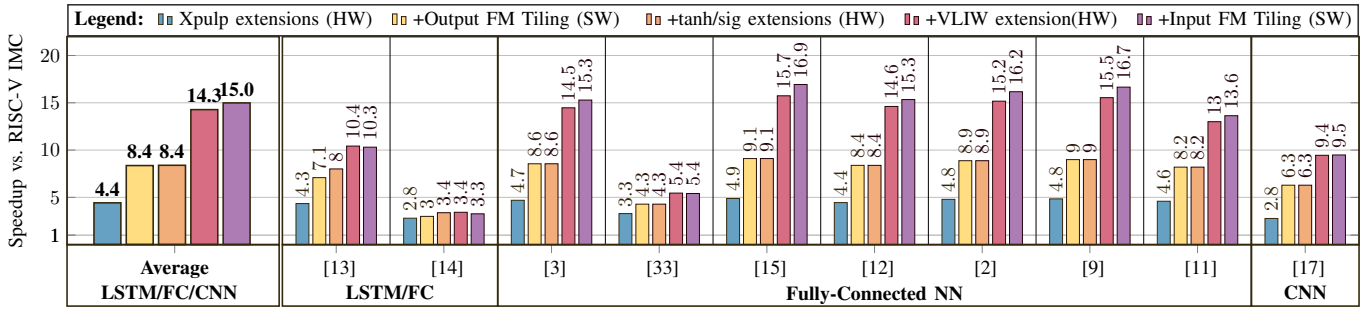}
 \vspace{-0.6cm}
\caption{Speedup with respect to the RISC-V IMC baseline implementation for a typical Neural Networks workload in RRM.}
\label{fig:relativecyclecount}
\ifdefined\optLevelTWO
\vspace{-3mm}
\fi
\end{figure*}

The results are shown in Tab.\,\ref{tab:rnninstructioncomparison}c and Fig.\,\ref{fig:relativecyclecount}, most of the networks execution cycles can be improved between 1.79\x{} \cite{Yu2017} and 1.87\x{} \cite{wang2018deep}, but small FMs suffer from high overhead and therefore less speedup (1.07\x{} \cite{Eisen} and 1.30\x{} \cite{Naparstek2017}). 

Overall, we obtain a speedup of 15$\times$ to the RISC-V IMC baseline thanks to: 4.4$\times $ using SIMD and HWL from the Xpulp extension, 1.9$\times$ with OFM tiling, 1.7$\times$ merging load and compute and 4.7\% with IFM tiling. 





\subsection{Tanh and Sigmoid Extension (HW)}

\ifdefined\specialtrialtoshrinkONE
\begin{figure*}[ht]
\begin{minipage}[b]{0.63\linewidth}
\centering
\includegraphics[width=\optLevelONEternary{1}{1}\textwidth]{\rnnfigures{extended_riscV-GRAY5.pdf}}
    \caption{RNN RISC-V Core with extensions to RI5CY core \cite{Gautschi2017} in blue and datapath for \texttt{pl.sdotpsp} instruction marked in bold.\vspace{0.8cm}}
    \label{fig:rnn_schematic}
\end{minipage}\hspace{2mm}
\begin{minipage}[b]{0.35\linewidth}
\centering
\def\speciallinebreak{}
\scalebox{0.72}{
\begin{tabular}{rc|c} %
    \begin{tabular}{@{}l@{}}\speciallinebreak1:\\
                            2:\\
                            3:\\
                            4:\\
                            5:\\
                            6:\\
                            7:\\
                            8:\\
                            9:\\
                            10:\\
                            11:\\
                            12:\\
                            13:\\
                            
                              \end{tabular} &
      \begin{tabular}{@{}l@{}}\speciallinebreak\\\\
                              \textbf{lp.setupi 0, 9, 32  // do \{} \\
                              \ \ \ \ lw rB, Imm(rBAddr!) \\
                              \ \ \ \ lw rA0, Imm(rAAddr0!) \\
                              \ \ \ \ lw rA1, Imm(rAAddr1!) \\
                              \ \ \ \ lw rA2, Imm(rAAddr2!) \\
                              \ \ \ \ lw rA3, Imm(rAAddr3!) \\
                              \ \ \ \ pv.sdotsp.h rD0, rA0, rB \\
                              \ \ \ \ pv.sdotsp.h rD1, rA1, rB \\
                              \ \ \ \ pv.sdotsp.h rD2, rA2, rB \\
                              \ \ \ \ pv.sdotsp.h rD3, rA3, rB \\
                              \textbf{//\}}
                              \end{tabular}   &  \begin{tabular}{@{}l@{}}\speciallinebreak
                              pl.sdotsp.h.0 r0, rA0, r0 \\
                              pl.sdotsp.h.1 r0, rA1, r0 \\
                              \textbf{lp.setupi 0, 5, 32  // do \{} \\
                              \ \ \ \ lw rB, Imm(rBAddr!) \\
                              \textbf{// bubble rB dependency}\\
                              \ \ \ \ pl.sdotsp.h.0 rD0, rA2, rB \\
                              \ \ \ \ pl.sdotsp.h.1 rD1, rA3, rB \\
                              \ \ \ \ pl.sdotsp.h.0 rD2, rA0, rB \\
                              \ \ \ \ pl.sdotsp.h.1 rD3, rA1, rB \\
                              \textbf{//\}}\\\\\\\\
                              \end{tabular} 
    \end{tabular}}  
    \vspace{4mm}
\captionof{table}{Assembly Code comparison with FM tiling only and with the \texttt{pl.sdotsp.h} instruction}
    \label{tab:assemblycompvliw}
\label{tab:asdf:image}
\end{minipage}
\end{figure*}

\else
\begin{figure*}[h!]
    
    \includegraphics[width=\optLevelONEternary{1}{0.75}\textwidth]{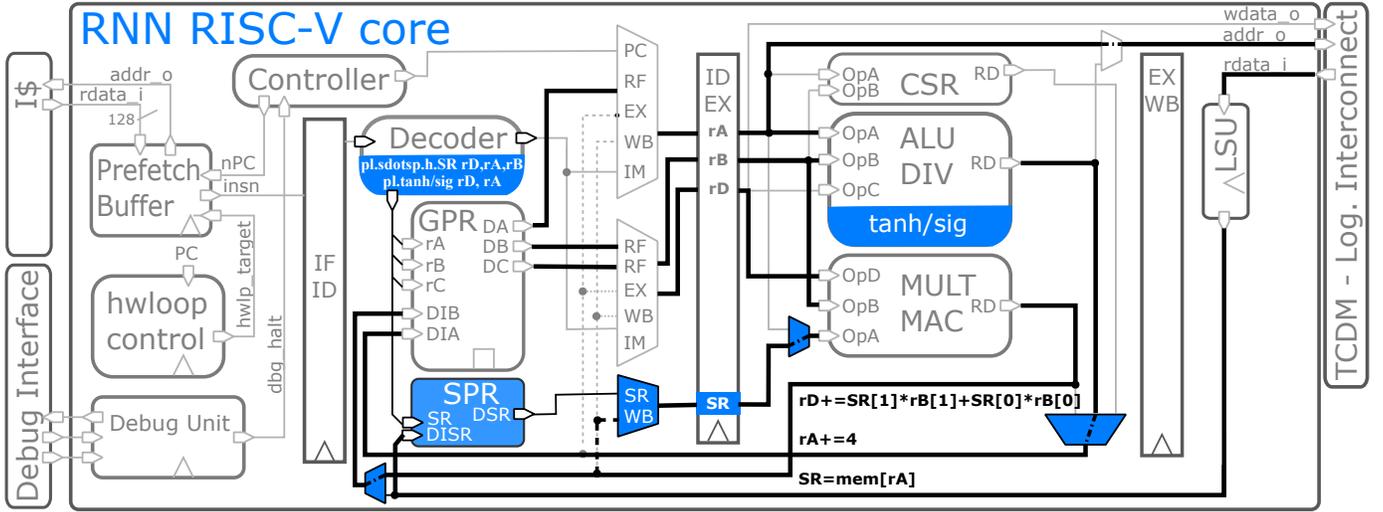}\scalebox{0.3}{}
    
    \ifdefined\optLevelONE
    \vspace{-0.2cm}
    \fi
    \caption{RNN RISC-V Core with extensions to RI5CY core \cite{Gautschi2017} in blue and datapath for \texttt{pl.sdotpsp} instruction marked in bold.}
    \label{fig:rnn_schematic}
\end{figure*}
\fi
Sigmoid and hyperbolic tangent are common activation functions in neural networks and used in LSTMs. The piecewise linear approximation technique can be implemented for these functions in SW with an increasing number of cycles to reach the required precision. This can be a major contribution to the overall calculation in LSTM-based networks. For example, the calculation of \texttt{tanh}/\texttt{sig} requires \calc{1}{6.2+4.1}\% in \cite{Challita2017} and \calc{1}{20.3+13.3}\% in \cite{Naparstek2017} of the overall computation cycles. We introduce two single-cycle instructions \texttt{pl.tanh rD, rA} and \texttt{pl.sig rD, rA} with the following useful properties:


\begin{enumerate}
\item They are continuous and smooth (i.e., derivatives are continuous, too); thus, the error is bound for a fixed interval in a Taylor series expansion even for degree one (i.e., $tanh(x_0+\epsilon)=tanh(x_0)+tanh'(x_0)\cdot \epsilon$).
\item The functions converge fast to either $0, 1$ or $-1$. Interpolation is needed only on this limited range of numbers.
\item Both functions are symmetric around 0 (i.e., $tanh(-x)=-tanh(x)$ and $sig(-x)=1-sig(x)$), thus just the positive number range needs to be interpolated and the negative range can be derived from the positive values.
\end{enumerate}
Alg. \ref{alg:tanhsigmoid} shows the pseudo-code that was used for the hardware implementation of the proposed interpolation. First, we chose the number of intervals of $M$ and the size of every interval $2^N$, whereas the interpolation range is $\pm M\cdot2^N$. For both functions $f\in \{tanh, sig\}$ two $M$-entry LUTs $\text{lut\_m}_f[\cdot]$ and $\text{lut\_q}_f[\cdot]$ are defined. Then the absolute value is calculated (line 2) and the index is calculated by a right shift of the absolute value by $N$ places, if the result is larger than $M$, it is considered to be in the convergence area and either $\{-1,0,1\}$ is returned. Otherwise, the value is calculated by linear approximation within the selected interval \textit{id} (line 8), sign inverted for negative values (line 9) and subtracted from 1 for negative values in the sigmoid case (l. 10).
\begin{algorithm}[h!]
  \caption{Pseudocode of the \texttt{sig} and \texttt{tanh} Interpolation}
  \label{alg:tanhsigmoid}
\begin{algorithmic}[1]
\REQUIRE{value $x$ and function $f\in{tanh(\cdot), sig(\cdot)}$, interval size $2^N$ and \#intervals M}
\STATE \STATE $|x|=\left\{
\begin{array}{ll}
-x & sgn(x)=1 \\
x & sgn(x)=-1
\end{array}
\right.$
\STATE $id=|x| >> N$
\IF{id$>$M}
    \RETURN $\left\{
\begin{array}{ll}
1 & sgn(x)=-1 \\
0 & sgn(x)=1, f=sig \\
-1 & sgn(x)=1, f=tanh \\
\end{array}
\right.$
\ELSE
    \STATE (m,q)=(lut\_m$_f$[id], lut\_q$_f$[id])
    \STATE $y=m|x|+q$
    \STATE $y=\left\{
\begin{array}{ll}
-y & sgn(x)=1 \\
y & sgn(x)=-1
\end{array}
\right.$
\RETURN $\left\{
\begin{array}{ll}
1-y & f=sig, \text{sgn}(x)=-1 \\
y & \text{else} \\
\end{array}
\right.$
\ENDIF
\end{algorithmic}
\end{algorithm}

\begin{figure}[h!]
    \centering
    \ifdefined\optLevelONE \vspace{-5mm}\fi
    \includegraphics[trim={0 0 0 5mm},clip, width=\optLevelONEternary{0.95}{0.85}\columnwidth]{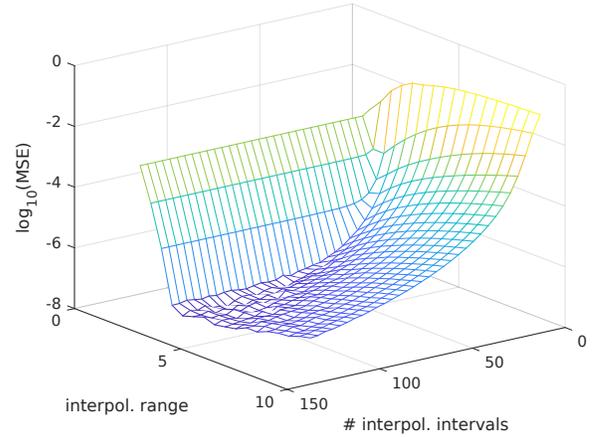}\ifdefined\optLevelONE \vspace{-4mm}\fi
    \caption{\texttt{tanh} Mean Square Error for different interpolation ranges and number of intervals and $Q_{3.12}$ quantization.}
    \label{fig:tanhmaxmse_dac}\ifdefined\optLevelONE \vspace{-5mm}\fi
\end{figure}

We evaluate the proposed piecewise linear approximation with different number of intervals $2^N$ and interpolation ranges, taking into account that fixed-point operations using the  $Q_{3.12}$ format are used. The result of this evaluation is illustrated in Fig.~\ref{fig:tanhmaxmse_dac}. For the actual implementation, we have selected an interpolation range of $[-4,4]$ and $2^5=32$ intervals, which produces an MSE of $9.81\cdot 10^{-7}$ and a maximum error of $\pm3.8\cdot 10^{-4}$ when compared to the full-precision hyperbolic tangent function. Evaluation of the quantized RNN benchmarks shows no deterioration of the end-to-end-error when replacing the activation function with our proposed interpolation, which is not surprising as Neural Networks are known to be robust against noise. This extensions reduce the cycle count from 51.2 to 44.5\,\textit{k}cycles within the LSTM networks \cite{Challita2017,Naparstek2017}, resulting in a \calc{1}{100-4454.6/51.174}\% improvement.

\subsection{Load and Compute VLIW instruction (HW)}

Analyzing the cycle counts in Tab.\,\ref{tab:rnninstructioncomparison}c, we can see that, the \texttt{lw!} and \texttt{pl.sdotsp.h} instructions dominate. By
introducing a new instruction, which combines these two within a single \texttt{pl.sdotsp.h} instruction which calculates a 16-bit packed SIMD sum-dot-product:\newline
\texttt{rD[31:0]+=rA[31:16]*rB[31:16]+ rA[15:0]*rB[15:0]}\\
but also loads data from the memory.
\ifdefined\specialtrialtoshrinkONE \else
\begin{table}[h]
\caption{Assembly Code comparison with FM tiling only and with the \texttt{pl.sdotsp.h} instruction}
    \label{tab:assemblycompvliw}
\vspace{-1mm}
    \ifdefined\arxiv \else\hspace{-13mm}\fi
    \ifdefined\optLevelONE
    \newcommand{\scalefactor}{0.9}
    \else
    \newcommand{\scalefactor}{1}
    \fi
    \scalebox{\scalefactor}{
    \begin{tabular}{lc|c} %
    \begin{tabular}{@{}l@{}}1:\\
                            2:\\
                            3:\\
                            4:\\
                            5:\\
                            6:\\
                            7:\\
                            8:\\
                            9:\\
                            10:\\
                            11:\\
                            12:\\
                            13:\\
                            
                              \end{tabular} &
      \begin{tabular}{@{}l@{}}\\\\
                              \textbf{lp.setupi 0, 9, 32  // do \{} \\
                              \ \ \ \ lw rB, Imm(rBAddr!) \\
                              \ \ \ \ lw rA0, Imm(rAAddr0!) \\
                              \ \ \ \ lw rA1, Imm(rAAddr1!) \\
                              \ \ \ \ lw rA2, Imm(rAAddr2!) \\
                              \ \ \ \ lw rA3, Imm(rAAddr3!) \\
                              \ \ \ \ pv.sdotsp.h rD0, rA0, rB \\
                              \ \ \ \ pv.sdotsp.h rD1, rA1, rB \\
                              \ \ \ \ pv.sdotsp.h rD2, rA2, rB \\
                              \ \ \ \ pv.sdotsp.h rD3, rA3, rB \textbf{//\}}\\
                              \end{tabular}   &  \begin{tabular}{@{}l@{}}
                              pl.sdotsp.h.0 r0, rA0, r0 \\
                              pl.sdotsp.h.1 r0, rA1, r0 \\
                              \textbf{lp.setupi 0, 5, 32  // do \{} \\
                              \ \ \ \ lw rB, Imm(rBAddr!) \\
                              \textbf{// bubble rB dependency}\\
                              \ \ \ \ pl.sdotsp.h.0 rD0, rA2, rB \\
                              \ \ \ \ pl.sdotsp.h.1 rD1, rA3, rB \\
                              \ \ \ \ pl.sdotsp.h.0 rD2, rA0, rB \\
                              \ \ \ \ pl.sdotsp.h.1 rD3, rA1, rB \\
                              \textbf{//\}}\\\\\\
                              \end{tabular} 
    \end{tabular}
    }
    
\ifdefined\optLevelONE \vspace{-6mm}\fi
\end{table}
\fi
Fig.~\ref{fig:rnn_schematic} shows the RI5CY core with the extended datapath of the \texttt{pl.sdotsp.h} instruction with the changes highlighted in colors and its active data paths in bold. \texttt{rA} contains the memory address, loaded from memory by the load/store unit LSU and is incremented for the next data access (i.e., next weight of the corresponding output channel). To avoid a 2-cycle latency, and thus unnecessary stalling, the data is stored in two special-purpose registers SPR and is written and read in an alternating way (using \texttt{pl.sdotsp.h.0} and \texttt{pl.sdotsp.h.1} instructions) from these two registers. The data from the SPR is multiplexed as 2nd operand \texttt{OpA} to the multiplier calculating the sum-dot-product. Data hazards are avoided by stalling the pipeline in case of missing grant from memory, exploiting exactly the same signals and control strategy used for standard load words.

Tab.~\ref{tab:assemblycompvliw} shows the assembly with output FM tiles of four with (right) and without (left) the extension. In lines 1-2, the SPRs are pre-loaded with the first two weights before the actual main loop. The input FM is loaded in line 4, which is used for the following MAC computation. As can be seen in Tab.\,\ref{tab:rnninstructioncomparison}d, the cycle count can effectively be reduced by 1.7$\times$.

Due to the latency of the load word and the dependency with the following instructions, a bubble is inserted in line 5. This can be further optimized by loading two input data (= four input channels), and the result calculated for all the output channels doubling the number of \texttt{pl.sdotsp.h} in the most inner-loop. However, the gains, as seen in Tab.\,\ref{tab:rnninstructioncomparison}e, are rather modest 1.05$\times$ (or 4.9\%) since loads and stores from the stack increase by \calc{1}{(2364+1487)/(1483+1354)}$\times$ as more registers are needed. 

Fig.~\ref{fig:relativecyclecount} shows the relative benefits to the RI5CY baseline compared to the output FM tiling, using the instruction extensions and the Input FM Tiling, where for most of the networks the input FM tiling has a positive effect, but few networks (i.e., NNs with small feature sizes) even need more cycles due to the increased stack operations.


\section{Core Implementation Results}

 
The extended RI5CY core was implemented in Globalfoundries 22\,nm FDX technology using an 8-track low-threshold (LVT) standard cell library and has been synthesized with Synopsys Design Compiler~18.06,  back-end flow has been done with Cadence Innovus~18.11 and power estimates are obtained by running gate-level simulations using Modelsim Questa v2019.1 with back-annotated delays from the final layout.
\ifdefined\removeforpaper \else Fig.~\ref{fig:piechart area} shows the area break-down of the RI5CY core without and with the new extensions. \fi
When compared to a standard RI5CY core (RV32-IMCXpulp), the new instructions result in a very small circuit area overhead of 2.3\,kGE (or \textcolor{black}{3.4} \% of the core area). 
Furthermore, the critical path of the core remains unchanged (between the load-store unit and the memory in the write-back stage), and the core operates at 380\,MHz at 0.65\,V at typical conditions at \ifdefined\optLevelTWO 25~\textdegree C \else room temperature \fi.
\ifdefined\removeforpaper \else
\begin{figure}[h!]
    \begin{tikzpicture}
    \pie[style=drop shadow, color={cieee4!50,cieee3!50,cieee2!50, cieee1!30, cieee1!50, cieee0!30, cieee0!70,cieee5!50, cieee6!50}, radius=1.2, text=pin]{2.107175829/lsu, 6.247486766/cs\_registers, 8.634318365/if\_stage, 12.93735684/{id\_stage w$/$o regs}, 21.48885554/{id\_stage$/$gpr}, 45.18747798/ex\_stage, 3.397328683/\textbf{RNN Ext.}}
\end{tikzpicture}
    \caption{Area Distribution of the block for the RNN ASIP}
    \label{fig:piechart area}
\end{figure}{}

\begin{table}[]
\begin{tabular}{lll}
                                               & RISCY & RNNSCY \\
ex\_stage\_i                                   & 0.47  & 1.04   \\
ex\_stage\_i/mult\_i                           & 0.34  & 0.73   \\
ex\_stage\_i/alu\_i                            & 0.06  & 0.14   \\
id\_stage\_i                                   & 0.69  & 0.85   \\
id\_stage\_i/registers\_i                      & 0.35  & 0.41   \\
id\_stage\_i/datapath                          & 0.31  & 0.41   \\
id\_stage\_i/decoder\_i                        & 0.01  & 0.01   \\
if\_stage\_i                                   & 0.29  & 0.35   \\
if\_stage\_i/prefetch\_32\_prefetch\_buffer\_i & 0.22  & 0.27   \\
load\_store\_unit\_i                           & 0.07  & 0.13   \\
other                                          & 0.22  & 0.24   \\
Total                                          & 1.74  & 2.61  
\end{tabular}
\caption{\textcolor{red}{fix table}}
\end{table}

\fi

When compared in the same core performing the RISC-V standard RV32-IMC instructions, when executing relevant RNN benchmarks, the enhanced core is \emph{on average} 15$\times$ faster. It performs 566\,MMAC/s (instead of 21\,MMAC/s). \revFIVE{When the core is using the extensions, the power consumption rises from 1.73\,mW to 2.61\,mW (\calc{0}{100*(2.61/1.73-1)}\% total increase). While the decoder contributes little more power ($\approx$5\textmu W), the higher power consumption is mainly due to the higher utilization of the compute units (ALU and MAC unit, i.e., 0.57\,mW/\calc{0}{100*(0.57/1.73)}\% of the total power), the increased GPR usage (0.16\,mW/\calc{0}{100*(0.16/1.73)}\%), and the higher use of the load-store unit (0.05\,mW/\calc{0}{100*(0.05/1.73)}\%). However, the overall energy efficiency at 218 GMAC/s/W shows a 10$\times$ improvement. }



\section{Conclusion}
We present the first RISC-V core design optimized for RRM applications using machine learning approaches based on RNNs. The core achieves an order-of-magnitude performance (15\x) and energy efficiency (10\x) improvements over the baseline RISC-V ISA on a wide range or RNN flavors used in RNN. These results are obtained thanks to a synergistic combination of software and hardware optimizations, which only marginally increase area cost and do not affect operating frequency.  It is essential to notice that the proposed optimization does not impact numerical precision. Hence labor-intensive quantization-aware retraining is not needed.  The enhanced RISC-V core achieves 566\,MMAC/s and 218\,GMAC/s/W (on 16-bit data types) in 22\,nm FDX technology at 0.65\,V, thereby providing a fully programmable and efficient open-source IP for future systems-on-chip for 5G Radio Resource Management. 

\section{Acknowledgments}
\ifdefined\anonymous
Hidden for double-blind review.
\else
This work was funded by Huawei Technologies Sweden AB. 
\ifdefined\optLevelTWO \else The authors would like to thank the PULP community for providing a comprehensive and open-source RISC-V platform.\fi
\fi







\ifdefined\arxiv \else
\fi
\todolater{replace the PULP-NN paper reference before }
\ifdefined\bibshortened

{\tiny\bibliography{sample-base,shortened}}

\else
\bibliography{sample-base,references_fixed}
\fi

\end{document}